**How to estimate heritability, a guide for genetic epidemiologists**

Ciarrah-Jane S Barry[1,2,*], Venexia M Walker[1,2,3], Rosa Cheesman[4], George Davey Smith[1,2], Tim T Morris[1,2], Neil M Davies[1,2,5]

[1] Medical Research Council Integrative Epidemiology Unit at the University of Bristol, Barley House, Oakfield Grove, BS8 2BN, United Kingdom

[2] Population Health Sciences, Bristol Medical School, University of Bristol, Barley House, Oakfield Grove, Bristol, BS8 2BN, United Kingdom

[3] Department of Surgery, University of Pennsylvania Perelman School of Medicine, Philadelphia, USA

[4] PROMENTA Research Center, Department of Psychology, University of Oslo, Oslo, Norway

[5]K.G. Jebsen Center for Genetic Epidemiology, Department of Public Health and Nursing, NTNU, Norwegian University of Science and Technology, Norway

* Corresponding author (email: ciarrah.barry@bristol.ac.uk)

Keywords: human genetics, epidemiologic methods, heritability, twin study, human genome, genome wide association study

ABSTRACT

Traditionally, heritability has been estimated using family-based methods such as twin studies. Advancements in molecular genomics have facilitated the development of alternative methods that utilise large samples of unrelated or related individuals. Yet, specific challenges persist in the estimation of heritability such as epistasis, assortative mating and indirect genetic effects. Here, we provide an overview of common methods applied in genetic epidemiology to estimate heritability i.e., the proportion of phenotypic variation explained by genetic variation. We provide a guide to key genetic concepts required to understand heritability estimation methods from family-based designs (twin and family studies), genomic designs based on unrelated individuals (LD score regression,

GREML), and family-based genomic designs (Sibling regression, GREML-KIN, Trio-GCTA, M-GCTA, RDR). For each method, we describe how heritability is estimated, the assumptions underlying its estimation, and discuss the implications when these assumptions are not met. We further discuss the benefits and limitations of estimating heritability within samples of unrelated individuals compared to samples of related individuals. Overall, this article is intended to help the reader determine the circumstances when each method would be appropriate and why.

## INTRODUCTION

Many human phenotypes are influenced by a complex mix of genetic and environmental factors. It is therefore important to comprehensively account for genetic influence when discerning how phenotypic variation arises. The broad role of genetics is commonly quantified using heritability; the proportion of phenotypic variation that can be statistically explained by genetic variation, see Appendix. There is evidence that most phenotypes are heritable, with heritability typically being higher for biological, e.g., eye colour, than social or behavioural traits, e.g., extraversion (1-5).

Historically, estimates for heritability originated from samples of related individuals. These methods capitalise on the known shared genetic variance between relatives, e.g., offspring share half of the genotype of each parent. Paths analysis has been widely implemented assuming a linear variance model. This separates the phenotype of interest into components; the additive genetic (A), common familial environment (C) and the environmental contribution unique to the individual (E), termed the ACE model (6). However, these methods require a range of assumptions about the cause of similarity within family pairs. The increasing availability of large samples of genotyped individuals stimulated the development of methods to estimate heritability within samples of unrelated individuals (7, 8). These methods capitalise on the large sample size and lack of environmental bias between unrelated individuals. However, these approaches are limited to capturing the additive component of the ACE model, thus a portion of heritability is missing from estimates (9). These techniques have more recently been extended to incorporate samples of related individuals, simultaneously accounting for the common environment and utilising the wealth of available genotyped data (10, 11).

Many methods exist to estimate heritability, which are dependent on different testable and untestable assumptions. These estimators can be influenced by demographic, familial and genomic factors such as population stratification, indirect genetic effects, assortative mating, linkage disequilibrium (LD) and epistasis (Box 1) (7, 12-15). The calculated estimate of heritability is specific to the population under study and may not be transferable across different populations across space or time. Heritability estimates generally fall into two categories: broad sense and narrow sense. *Broad sense heritability* is the proportion of phenotypic variation that is statistically explained by total genetic variation, including dominance and epistasis (see Box 1). *Narrow sense heritability* refers to phenotypic variation explained by additive genetic variation only (13, 16).

Here, we describe methods for estimating heritability, the assumptions upon which they rest, and the biases that may be introduced to each method under specific circumstances. We have selected approaches that are widely implemented in the literature and group them into three sections (Figure 1): 1) family-based designs, 2) genomic designs based on unrelated individuals and 3) family-based genomic designs. We review these methods and approaches to estimate heritability and discuss their respective benefits and limitations.

Figure 1: A Flowchart Detailing the Outline of the Review

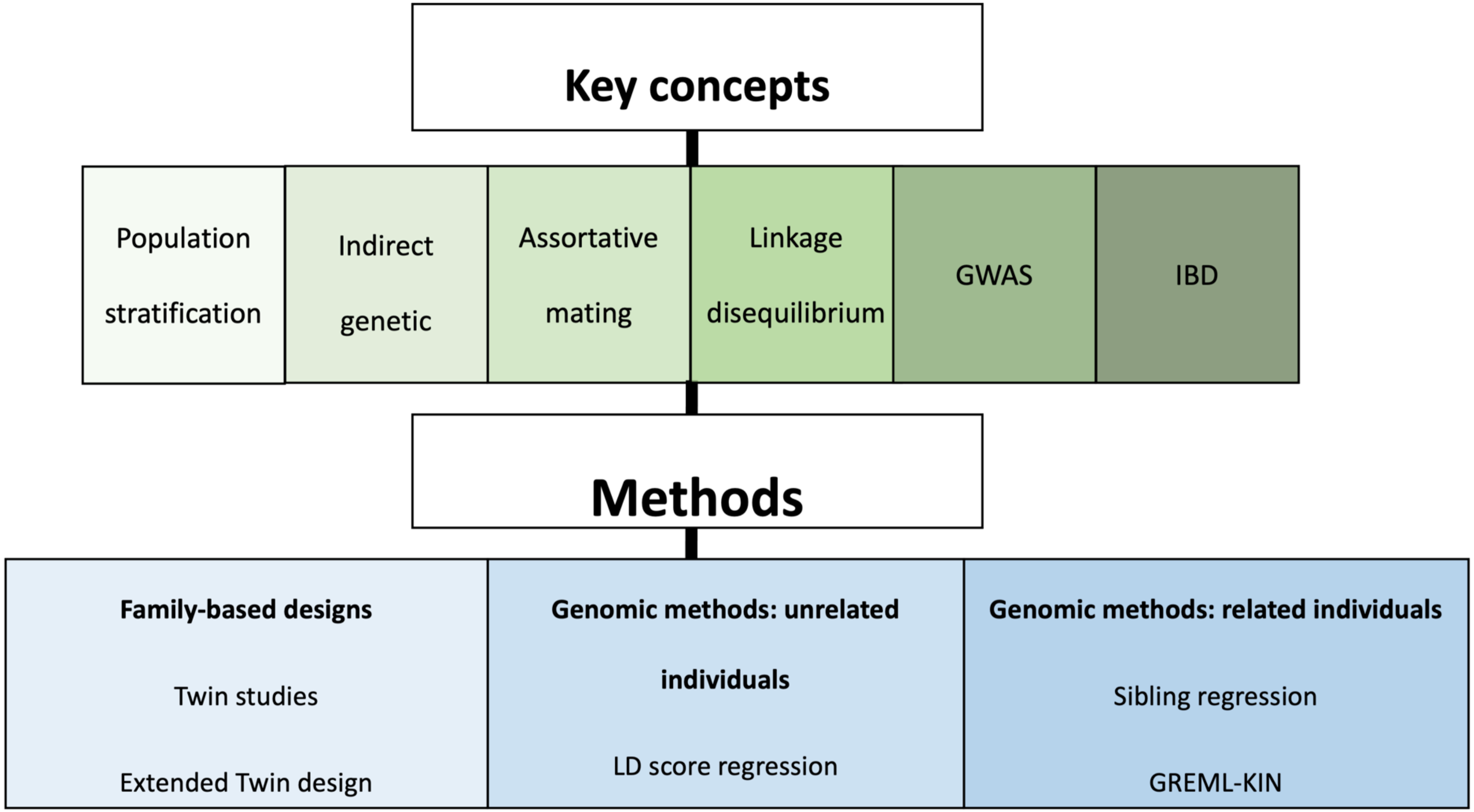

Box 1: Key Genetic Concepts That may be Useful to Understand the Methods Described Throughout This Review

Key concepts

- *We define genetic variants with minor allele frequency (MAF) > 5% to be common variants, SNPs with MAF 1-5% to be low-frequency variants, and genetic variants with MAF < 1% to be rare variants (13, 14).*

- *Epistasis* is the effect of interaction between genes at multiple locations (17, 18). Epistasis alters the variance of a phenotype (19, 20). The effects between one variant and a variant in a different stratum of a single parameter can have effects in opposing directions. Hence the genetic effects of the parameter may not be detectable if one gene is counteracting the expression of the other (21). There is not yet a consensus on the importance of epistasis. Some argue epistatic effects play a minor role in phenotypic expression relative to additive effects (22-25), yet others propose epistasis may greatly impact upon estimates (26-29), which would consequently influence estimates of heritability obtained via methods that do not account for it (20, 30, 31). Current studies appear to only detect interactions with very large effects; therefore, it is possible many more interactions exist but are undetectable (1, 32-35).

- *Dominance* refers to allelic interactions within a gene (36). Dominance variation has been found to contribute little to SNP based estimates of heritability (37).

- *Population stratification* refers to differences in allele frequencies across groups within a population (11, 38, 39). These allele frequency differences can arise from non-random mating and geographic separation. Members of isolated populations experience random changes in allele frequency over time (genetic drift), which can lead to observable differences in allele frequency over many generations (40). It is widely recognized that population stratification can inflate estimates of heritability, although this is method dependent. For example, in some within-family genetic methods the estimate for heritability is not affected method (11, 14, 41-43).

- *Direct genetic effects* are the effects of an individual's own genotype on their own phenotype (44).

- *Indirect parental genetic effects* occur when parental genotype affects offspring phenotype through its expression in the parental phenotype (12, 45). For example, parents with more education-associated alleles may procure more books for their households than parents with fewer education associated alleles (the phenotypic expression of the parents' genetic variation). This may in turn positively influence their offspring's reading ability or knowledge, resulting in higher offspring educational outcomes. Indirect parental genetic effects can inflate heritability estimated within the offspring generation in certain situations. Indirect genetic effects are a similar concept to genetic nurture, dynastic effects, and, within twin studies literature, passive gene-environment correlation (46, 47).

- *Assortative mating* occurs when people choose mates non-randomly with respect to their phenotypes. Assortative mating can be single phenotype (e.g., education to education) or cross phenotype (e.g., education to height). Assortative mating can induce correlations between phenotypes and genetic variation for different phenotypes (12, 13, 48). Thus, when assortative mating is present, heritable traits become non-randomly distributed as partners will be more, or less, genetically similar than expected by chance which may invalidate key assumptions required by the method.

- *Linkage disequilibrium (LD)* is the non-random association of alleles at different loci within a population. Loci are in LD when the frequency of association of their different alleles is higher or lower than would be expected by chance (49). Genome-wide LD conveys a population's history, for example geographic subdivision. Genomic region LD reflects the gene-frequency evolution, including mutation, natural selection, and gene conversion (49, 50). LD can either be local, i.e. correlations between genetic variants that are in similar positions in the genome and are inherited in a block, or non-local in which more widely spaced genetic variants are correlated (51).

- *Identify-by-descent (IBD)* is a fundamental concept defining genetic relatedness between individuals (52). Two or more individuals who have the same genotype at a given point in the genome are identity-by-state (IBS). Further, individuals are IBD if they are IBS and have inherited this variation from a common ancestor. Therefore, the difference between IBS and IBD stems from knowledge of the relationship between the sequence and its origin (53), illustrated in figure B1. The covariance between the percentage of the genome shared between related pairs in a population (e.g., siblings or cousins) and their phenotypic similarity can be estimated using IBD segments discovered by dense genotyping, which covers a wide range of SNPs in a population (30, 54, 55). Heritability estimation methods that use IBD as a measure of relatedness have been developed (56).

Figure B1: Diagrams Demonstrating the Concepts of IBD (left) and IBS (right) Using Parental-Offspring Genotypes (57).

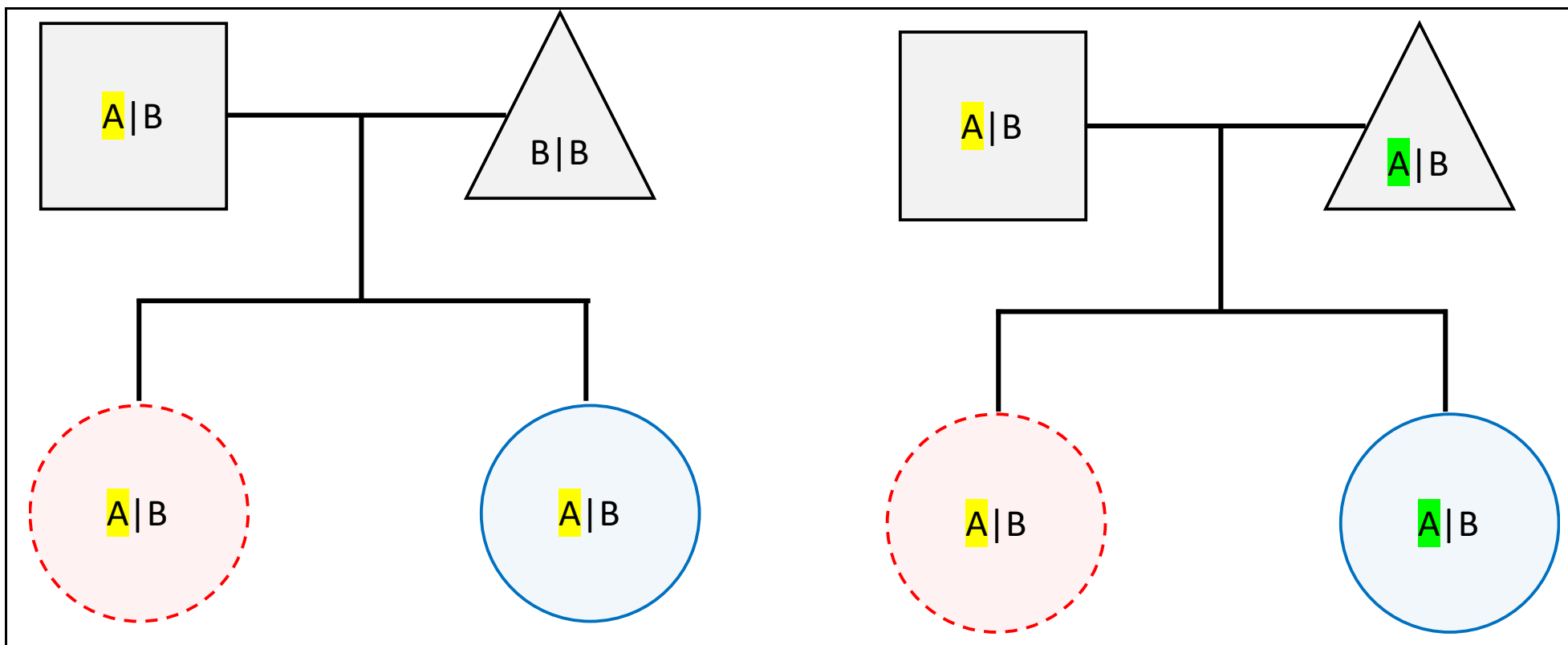

- *Genome-wide association studies (GWAS)* estimate millions of SNP-phenotype associations in large samples, vastly accelerating the study of the genetic architecture of complex phenotypes (58). Individual SNPs generally explain a small proportion of phenotypic variability as the magnitude of association is small. Hence, associations for many SNPs in a GWAS may not reach the genome-wide significance threshold ($p<5\times10^{-8}$) in available samples (59-61). This strict p-value threshold is required because of multiple testing and to reduce false positive discoveries (62). Most phenotypes and diseases are influenced by many genetic and environmental factors (known as the omnigenic model of inheritance) (63, 64). Many genomic methods to estimate heritability are reliant upon data obtained from GWAS (65).

## FAMILY-BASED DESIGNS

Family-based designs are methods that use samples of closely related individuals – for instance, siblings or parents and offspring. Family-based designs can be used to estimate heritability even if molecular genetic data is unavailable.

### Twin studies

Twin studies examine phenotypic differences between monozygotic (MZ) twins, who are genetically identical at conception, and dizygotic (DZ) twins, who on average share 50% of their segregating genetic variation (66), the genetic variation that results in individual differences between the pair (66). If MZ twins are more phenotypically similar than DZ twins, then this reflects genetic influence, because MZ twins are more genetically similar than DZ twins (30, 67, 68). The classic twin design is also known as the ACE model. Here, phenotypic variance is partitioned into attributable to additive genetic (A), common environmental (C) and non-shared environmental (E) variance components. The ACE model assumes that dominance genetic variance is zero. Hence, an estimate of narrow-sense heritability may be determined in the absence of common environmental effects on the phenotype (69). To the extent that MZ and DZ twins experience parental indirect genetic effects similarly, these effects will be included within the common environment (C) component (70). Twin studies may also be used to study gene-environment interactions (71). Through implementation of an ADE model, in which variance components are modelled as additive (A), dominance and epistasis (D) and non-shared environmental effects, estimated heritability may be partitioned into additive and dominance variance components.

Twin studies require several assumptions. First, it is assumed the shared environment makes an equal contribution to the phenotype of interest across both MZ and DZ twin pairs, termed the equal environment assumption (EEA) (11, 16, 72, 73). If the EEA is invalid, estimates of heritability are likely to be inflated because differing environments would be mistakenly attributed to differences in genetic variation (9, 72, 74, 75). The validity of the EEA has been tested within the literature through a comparison of trait similarity as a function of independently assessed zygosity and family-perceived zygosity, with respect to psychiatric disorders. Consistent with previous findings, little evidence was found to suggest perceived zygosity influences twin resemblance. Hence, differing expectations contingent on zygosity were not determined to substantially bias twin study findings and violation of the EEA will have little effect (75). Further, simulations have demonstrated the impact of a violation of the EEA is likely to be modest, ranging between a 7-14% reduction in heritability estimate (76). A second assumption is that twins are generalisable to the general population with regards to the phenotype of interest. The validity of this assumption has been demonstrated in multiple studies (67, 77, 78). The third assumption is that random mating occurs within the population (13, 66, 74, 76, 77, 79-82). Finally, studies have suggested that the confounding role of environmental similarity on genetic factors and outcomes is limited, Figure 2 (83, 84). If all these assumptions hold, then a comparison of the phenotypic correlations of MZ and DZ twins can reliably estimate heritability (66, 67, 74).

Figure 2: A Directed Acyclic Graph Demonstrating the Concept of Confounding.

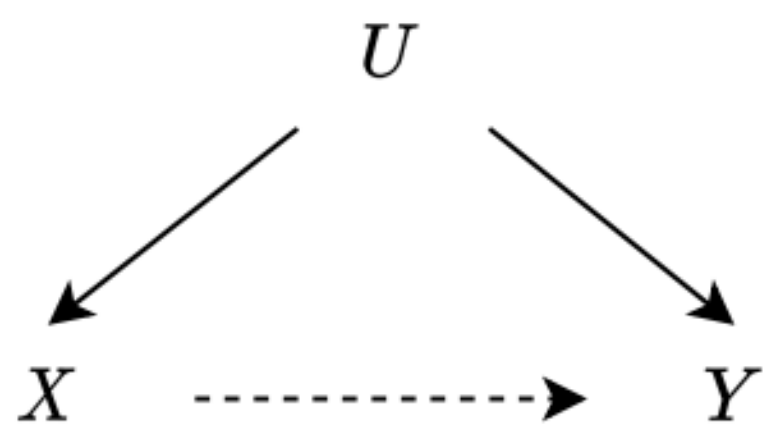

Extended twin designs

Extended twin designs model the similarity between other relatives alongside the twins. This enables additional parameters to be estimated. For example, it is possible to determine the twin-specific and non-parental components of the shared environment, alongside the additive and non-additive components of genetic variance (85, 86). Additionally, the effect of assortative mating may be accounted for through the inclusion of spouses (87).

GENOMIC METHODS: UNRELATED INDIVIDUALS

The mass characterisation of the human genome has generated a rich, vast resource for genetics research (88). Measured genome methods refer to statistical methods that are applied to molecular genetic (including whole genome) data, which has either been directly measured or imputed using reference panels. Some measured genome methods focus on the estimation of “SNP heritability”, which is a special case of narrow sense heritability estimated from measured SNPs.

LD Score regression

Linkage disequilibrium score regression (here LDSR, but sometimes referred to as LDSC) is a regression-based method that can separate genetic and confounding effects and estimate SNP-heritability (89). SNPs with high LD scores have a greater likelihood to tag causal SNPs, thus have a higher magnitude of association on average than SNPs with low LD scores (89-91). An LD score is created within a population reference panel for each SNP to represent the amount of tagged genetic variation explained by the SNP, see Appendix. The reference panel accounts for LD structure within a given population, hence, population stratification bias can be separated from genuine polygenicity. Summary statistics from multiple SNPs are regressed on the LD score of each SNP of interest, with the estimated intercept quantifying the heritability (90). Here the slope of the LDSR is rescaled to estimate the heritability explained by all SNPs that were used to estimate the LD scores (89). LDSR enables inflated GWAS test statistics to be differentiated from confounding bias. SNPs may tag both individual large and multiple weak effects, in contrast to other genome-wide methods (92). GWAS summary statistic based LDSR is more efficient than LDSR or other methods that use measured genotype data in a single sample, see Appendix. However, LDSR estimates are less accurate where fewer SNPs are available (93).

LDSR is dependent on several assumptions. First, it assumes the variance explained per SNP is assumed to be uncorrelated with LD score. Therefore, rare SNPs are assumed to have larger effect sizes than common SNPs. However, this may not hold in all circumstances, such as studies of phenotypes with correlated LD score and MAF (89). Second, the target sample of interest is assumed to be well matched to the LD reference panel (89). If this does not hold the accuracy of estimates decreases as the genetic heterogeneity discrepancy between the reference and sample of interest increases (91, 93). However, population stratification resulting from genetic drift does not correlate with LD and cannot be distinguished by LDSR.

GREML

Genomic relatedness restricted maximum likelihood estimation (GREML) estimates SNP heritability from measured genomic data. Importantly, the method uses unrelated individuals, who vary randomly in genetic similarity by chance. The method capitalises on the logic that in a population of unrelated people, more genetically related individuals are more phenotypically similar. Practically, this method is implemented by constructing a genetic relationship matrix (GRM) capturing the degree of relatedness between every pair of individuals at every SNP location, see Appendix (55, 73). The extent that the genetic matrix predicts phenotypic similarity reflects the degree of heritability.

First, it is assumed that all genetic effects are direct (i.e., biological effects in the offspring), even though transmitted variants can also influence phenotypes indirectly via the parents. If indirect genetic effects exist, these will be attributed to direct effects as illustrated in Figure 3. Consequently GREML overestimates the contribution of direct genetic effects and inflates heritability estimates of certain phenotypes in the presence of indirect effects (11).

Second, it is assumed that individuals do not share environmental influences (60). The restriction to unrelated individuals (typical threshold <0.025) limits confounding by common environmental effects alongside reducing contamination of non-additive genetic effects (7, 65, 73, 94, 95). Additionally, GREML assumes that there are no epistatic effects, and all estimated genetic effects are assumed to be additive. Third, it is assumed GREML estimates only capture the direct, additive effects of common SNPs, and not rare genetic effects or non-additive genetic effects. SNP-based methods may therefore be used to validate estimates derived from family-based methods requiring alternative assumptions (8). Thus, as with other methods using measured SNP data, estimates obtained via GREML can be considered to provide a lower bound of heritability (96). However, this lower bound should be used with caution as there are potential sources of inflation within GREML estimates that would not occur in family-based design studies. Fourth, random mating is assumed, meaning that GREML heritability estimates may be biased in the presence of assortative mating because of directional LD between SNPs, see Appendix (73, 97). Non-random selection increases, or decreases, additive genetic variation through the gametic phase disequilibrium. This means gene alleles are no longer randomly associated as SNPs occur with a frequency greater, or less, than the product of the frequency of the two relevant alleles, as would be expected through random mating (98). For example, GREML likely overestimates the contribution of genetic variation for phenotypes that are impacted by assortative mating, e.g., intelligence (99). Fifth, strong assumptions about genetic architecture, i.e., the characteristics of genetic variation that cause heritable phenotypic variability, are necessary (100). GREML assumes that SNP genotypes are standardised, with normally distributed effects that are independent of LD. Thus, SNPs with lower MAF are implicitly assumed to have a larger per-allele effect (7). Finally, it is assumed that the true genetic effect variance-covariance structure between pairs of individuals is known. In practice, this is an estimate derived from the GRM rather than the true genetic effect variance-covariance structure (7). In samples of unrelated individuals, this estimate is reasonable as genetic and environmental similarity are assumed to be independent, however this may be violated in the twin study setting.

In the presence of population stratification, standard GREML methods are likely to overestimate heritability. Specific GREML population-based methods include adjustment for the first principal components of the estimated kinship matrix into the model with fixed effects to account for population stratification (42, 64, 65, 73, 101). However, these are not thought to comprehensively control for the bias induced by population stratification, so some bias remains (14, 16, 41, 102). Additionally, GREML methods are sensitive to LD. GREML heritability estimates reflect LD between SNPs and unmeasured genetic variation. As a result, estimates using data from sparse SNP arrays with relatively few SNPs will be smaller than estimates from data with a denser array (7). Furthermore, common SNPs may not tag less common SNPs well.

The maximum LD correlation will decrease as the difference in MAF increases, thus, heritability estimates will be biased when SNPs are located in genomic regions with different LD properties to the rest of the genome (e.g., if regions of high LD, such as the human leukocyte antigen (HLA), the complex of genes coding the proteins that regulate the human immune system, are particularly important for a phenotype) (103).

In addition, GREML is highly sensitive to uneven LD, over- and under-estimating heritability in areas of high and low LD respectively. This is because correlation between SNPs distorts their estimated contribution to heritability. LD adjusted kinship (LDAK) has been proposed to reduce bias and increase precision in these circumstances. This calculates a SNP weight in relation to how well it is tagged by genomic neighbours (104). Weighting estimates were found to largely eliminate bias relative to estimates calculated using standard kinship matrix, however effects may be overstated. Alternative methods address LD heterogeneity but are computationally intensive (94, 105).

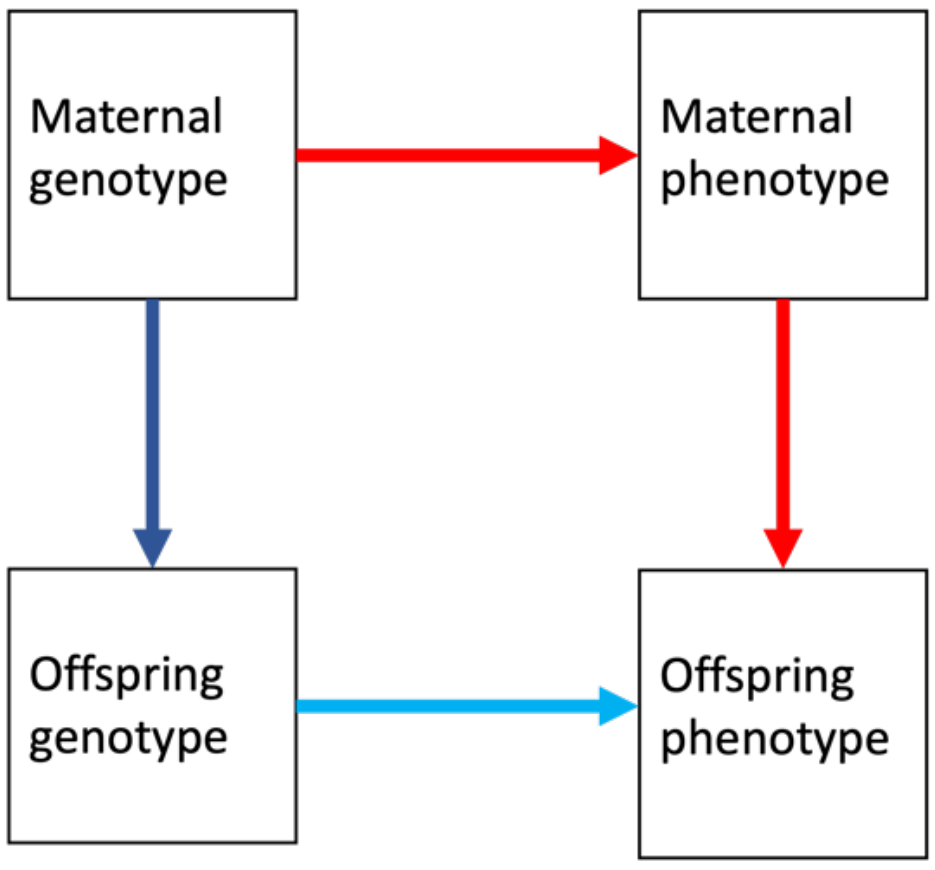

Figure 3: An Illustration Demonstrating Genetic Effects Between a Mother and Offspring.

GREML-LDMS

SNP based estimates of heritability can be biased when LD differs between causal variants and other variants. However, it has been demonstrated that this bias may be reduced by stratifying GREML heritability estimates based on a joint model of LD and MAF, termed GREML-LDMS (7). This is a multi-component approach in which SNPs are imputed by their MAF and regional LD. LD scores are calculated for each segment, and from this SNPs are stratified into quartiles. A GRM is then calculated from the stratified SNPs for each sample and a REML analysis is carried out using multiple GRMs (106). Simulations have demonstrated GREML-LDMS heritability estimates to be less biased compared to other GREML based estimation methods, though they are less precise (94, 105).

GREML-LDMS estimates require the following set of assumptions regarding the GRM to hold. First, that there is no gene-environment correlation. Second, that SNP effects are normally distributed, independent of LD and inversely proportional to MAF. Third, a much larger sample size (>30,000) is required for GREML-LDMS than the standard GREML approach (94).

## GENOMIC-METHODS: RELATED INDIVIDUALS

A shared limitation of the genomic methods applied to unrelated individuals is an inability to account for environmental confounding (Figure 2). A variety of methods have been developed to enable the implementation of genomic methods in large samples of related individuals.

### Sibling regression

An approach that can be used to estimate heritability within sibling pairs is derived from identity-by-descent (IBD) (107). IBD can be used to estimate narrow sense heritability and can be implemented in samples of individuals who are not closely related (56, 68, 107-109). Linear mixed modelling using IBD kinship matrices can be used to estimate the additive genetic effect (30). On average, siblings are 50% genetically similar (IBD). Variation around this 50% average exists due to random segregations at conception, which are approximately independent of most environmental effects (including indirect genetic effects) (11, 110). Hence, within sibling pairs, if genetic variation affects a phenotype, then siblings who are more similar and share more of their genomes IBD should be more phenotypically similar (111). Note, estimates obtained from siblings capture less of the genetic variance than estimates obtained by twin studies. Heritability estimated obtained via sibling regression do not include indirect genetic effects as this method is limited to narrow-sense heritability (111).

Statistical methods using sibling regression to estimate heritability are dependent on the following assumptions. First, estimates must also account for regions of LD (112, 113). Second, sibling regression assumes that estimates of percentage shared IBD between siblings across the entire genome are proportional to the number of causal additive SNPs between siblings for the trait of interest (107). Third, siblings who have inherited an IBD segment from a common ancestor have identical genetic segments in that region, hence estimates include the rare variant effects in that region (except de-novo mutation and other variant introducing events) (114, 115). Fourth, the additive genetic covariance between relatives is proportional to the proportion of the genome that is shared IBD (107). Finally, estimates are reliant on the assumption of random mating. However, sibling regression estimates may be inflated by assortative mating as the induced phenotypic correlation increases the genetic and phenotypic variation in the population, alongside relative pairwise phenotypic covariances. Thus, the common environmental effect between individuals could be overestimated and result in inflated heritability estimates, unless assortative mating is accounted for (13, 72).

GREML-KIN

The following four methods GREML-KIN, M-GCTA, Trio-GCTA and relatedness disequilibrium regression (RDR) are extensions to GREML. GREML-KIN estimates heritability whilst controlling for the shared environment through the inclusion of relatives. M-GCTA, Trio-GCTA and RDR are similar approaches as the phenotypic variance of the offspring is decomposed into direct and indirect genetic effects through the inclusion of one or both parental genotypes.

GREML-KIN is a modification of GREML that enables the estimation of additional genetic and shared environmental effects amongst family members (116) . Specifically, estimates of heritability can be segregated into the variance corresponding to family-specific genetic effects, and genetic effects at the population level. Two GRMs are included, one capturing the genetic relatedness of family members, and the other capturing genetic similarity among unrelated individuals. Matrices indexing environmental sharing are included to control for the fact that familial genetic similarity is correlated with environmental similarity (116). Since LD is stronger between more related individuals (compared with less related individuals), the family relatedness matrix in GREML-KIN may capture effects of variation not tagged by common SNPs. However, relative to other methods using the same sample size, smaller effects are harder to detect in GREML-KIN. For example, if the GRM contributes less that 5% of the overall phenotypic variance, only the major contributing components to the trait will be estimated reliably (116, 117). Further, GREML-KIN requires a sample comprised of a sufficient number of participants with ranging relatedness. Whether a dataset contains sufficient participants for GREML-KIN can be determined via simulations (117).

Whilst GREML-KIN can distinguish heritability and shared familial environmental effects, it is not a within family-specific method and therefore heritability estimates may still be confounded by indirect genetic effects.

M-GCTA

Maternal-GCTA (M-GCTA) estimates the proportion of offspring phenotypic variation that can be explained by both maternal and offspring genetic variation (118). The covariance between the offspring genetic effect on offspring phenotype, and the maternal genetic effect on maternal phenotype, is used to estimate the contribution of the indirect maternal effect to the offspring phenotype of interest (10). Therefore, within M-GCTA the maternal effects include the indirect effects of the maternally provided environment (including intrauterine effects), which would typically bias standard GREML estimates (10, 118). M-GCTA requires genetic information from large samples of mother-offspring (or father-offspring) pairs to model indirect genetic effects (10, 119), which may limit sample size compared to GREML in unrelated individuals. A further limitation to the method is the inability to simultaneously account for both parental genotypes, thus estimates may be biased by the genetic effects of the parent omitted in the model (10).

Most assumptions of standard GREML are common to M-GCTA, however M-GCTA does not rest upon assumptions of sample relatedness. M-GCTA additionally assumes that variance due to indirect maternal effects and correlations between direct and indirect maternal effects are assumed to be non-zero (10, 65).

Trio-GCTA

Trio-GCTA is very similar to M-GCTA however it includes genomic information from both parents (120). The presence of indirect effects from both parents can be tested empirically with samples of mother-father-offspring trios. The calculation of the direct and indirect genetic effects can apply to any member of the trio, although interpretation of parameters is specific to the individual of interest (120). It has been stated it is unclear how assortative mating could influence estimates, however we note the direct effect estimates from Trio-GCTA are robust to both population stratification and assortative mating (13, 120, 121) .The main limitation of trio-GCTA is the requirement for large samples of genotyped parent offspring trios. Most assumptions of standard GREML are common to Trio-GCTA, excluding the assumption of sample unrelatedness and assumptions about the structure of LD within the sample. Violation of the random mating assumption has an undetermined impact on inferences from Trio-GCTA at present (99, 120). In addition, Trio-GCTA has strict requirements about the distribution of the genetic and residual effects, and assumes they are independent, identically distributed and follow a multivariate normal distribution.

Relatedness disequilibrium regression

RDR estimates the influence of the indirect genetic parental effect on the offspring on top of the direct genetic effect, akin to separating out unmeasurable heritable phenotypes (11, 107). RDR differentiates the direct genetic effects from the indirect based on the random segregation during meiosis. Specifically, it allows us to decompose the elements of the phenotypic variation into direct genetic and other effects (e.g., environmental) and exploit the independence of offspring genotype and environment conditional on parental genotype (122). This is likely to hold within pairs of individuals who are not related by direct descent as they are unlikely to affect one another's residual environments. Here consistency is not affected by environmental confounding. Note, population stratification may bias estimates of parental indirect genetic effects. An alternative was to apply RDR is through the use of IBS segments, however the estimates produced are very similar (11).

Specific assumptions must be met within this method. First, direct genetic effects are assumed to be additive, e.g., no epistasis occurs. It may be possible to incorporate non-additive genetic effects within estimates, but consideration must be given to the corresponding non-additive associations between parental genotype and environment to ensure residual environmental effects are uncorrelated with the non-additive effects (11). Second, we have random mating, with analogous reasoning to previous methods. Third, for consistency of the estimator it is necessary that the trios are independent. This prevents segregation events within the individual pairs' parents becoming dependent on each other, which could induce bias comparable to the proportion of direct descendent related pairs (11).

Box 2: A Summary of the Strengths and Limitations of Each Discussed Method

| Method | Strengths | Limitations |
| --- | --- | --- |
| Twin studies | Estimates not substantially affected by violation of the EEA assumption, provides an upper bound estimate of heritability, incorporating the effects of rare SNPs, within pair effects not impacted by population stratification. | Estimates may be biased upwards due to shared environmental effects interacting with additive genetic effects, cannot determine the effect of epistatic interactions, estimates of the common environmental effects inflated by assortative mating. |
| Extended twin studies | May differentiate the additive and non-additive components of genetic variance, in addition to the effects of assortative mating. | Cannot account for inflation in estimates of non-additive effects |
| Sibling regression | Can provide an estimate of narrow sense heritability, no additional assumptions about the distribution of SNP effects, incorporates the effects of rare SNPs into heritability estimates, robust to genotyping errors and some missingness, partially accounts for population stratification. | Large sample size required for precise results due to the small standard deviation of IBD shared between siblings e.g., Visscher et al determined the average proportion of the genome-shared IBD within sibling pairs to have a standard deviation of 0.036 within their sample (11, 17, 107), the estimate is relative to a chosen reference population. |
| LD score regression | Computationally efficient as it only requires summary level data, partially accounts for population structure, see Appendix, precision increases with a greater number of SNPs, if the variance explained by each SNP is uncorrelated with LD score estimates are unbiased (contingent on other assumptions), possible to implement using free online tool (90). | Rare SNPs are assumed to have a larger effect, accuracy of estimates is dependent on a well-matched population panel, cannot estimate total heritability, a large sample is required to have reasonable power when detecting SNPs with lower heritability. |
| GREML | Provides a 'lower bound' estimate for heritability, possible to implement using widely available software, partially accounts for population stratification, can be implemented in large scale biobanks of unrelated individuals. | Cannot distinguish direct and indirect genetic effects which may inflate estimates, samples restricted to genotyped individuals, estimates inflated by assortative mating, additional assumptions about SNP effect sizes required, not suited to estimate the contribution of rare SNPs, estimates are highly sensitive to LD. |
| GREML-LDMS | Reduced bias resulting from LD compared to GREML-SC | Less precise estimates relative to other GREML methods |
| GREML-KIN | estimate heritability along with shared environmental effects coming from siblings, parents and spouses. | Requires samples with ranging relatedness, effect estimates may still be confounded by unmodelled shared environmental factors, greater power needed to detect smaller effects. |
| Trio-GCTA | Able to account for indirect genetic effects within heritability estimates, assumptions about the structure of LD are not necessary, partially accounts for population stratification. | Requires genotyped parent-offspring trios, additional assumptions about the distribution of genetic and residual effects are necessary, assortative mating and epistatic effects will bias estimates. |
| M-GCTA | Able to account for indirect genetic effects within heritability estimates, possible to implement using freely available software. | Requires genotyped maternal or paternal – offspring pairs, may be biased by assortative mating, large sample size required, cannot simultaneously account for both parental genotypes. |
| RDR | Environmental effects are not included in heritability estimates, arguably less constrictive assumptions are required than for alternative genome-wide family methods, such as the distribution of SNP effects, simulations have demonstrated population stratification can be mostly accounted for (11), a greater proportion of variance from rare SNPs may be captured relative to other genome-wide methods. | Epistasis and assortative mating will distort estimates, samples cannot contain individuals related by direct descent, population stratification may modify the assumed IBD sharing relationship. |

## CONCLUSIONS

A range of methods are available to estimate heritability from samples of closely or more distantly related individuals. However, it is important to consider the suitability of the estimator within the context of the study and the sample used for estimation. There are many factors that may impact on the accuracy and stability of heritability estimates such as the measurement of phenotypes or the representativeness of the sample relative to the wider population. An inherent advantage of studies involving families relative to studies of unrelated individuals, is the control of bias due to population stratification. Studying individuals from the same family ensures the same source population, thus a homogenous ancestry (12, 111, 123). The greater availability of large sample GWAS data has refocused efforts to estimate additive genetic heritability, generally through SNP based methods. Yet, individual SNPs contribute to a very small proportion of phenotypic variability relative to estimates from family-based methods. Many of the methods here have been developed to reduce the impact of known population phenomena and account for rare SNPs and epistasis, thus incorporating greater genetic information into estimates of heritability. Therefore, methods to estimate heritability should be selected with consideration of the sample of interest, and estimates should be interpreted with necessary caution. Triangulation of heritability estimates across different methods with different assumptions within samples is likely to provide the most robust evidence for the heritability of phenotypes, with consideration for any expected estimand differences

## AUTHOR AFFILIATIONS

Medical Research Council Integrative Epidemiology Unit at the University of Bristol, Barley House, Oakfield Grove, United Kingdom; Population Health Sciences, Bristol Medical School, University of Bristol, Barley House, Oakfield Grove, Bristol, United Kingdom (CJSB, VMW, GDS, TTM, NMD), Department of Surgery, University of Pennsylvania Perelman School of Medicine, Philadelphia, USA (VMW), PROMENTA Research Center, Department of Psychology, University of Oslo, Oslo, Norway (RC), K.G. Jebsen Center for Genetic Epidemiology, Department of Public Health and Nursing, NTNU, Norwegian University of Science and Technology, Norway (NMD).

## AUTHORSHIP

CJSB prepared the original draft, revised, and edited it for submission. VW, RC, GDS, TM and NMD supervised, reviewed, and edited the manuscript. All authors have approved the final version.

## FUNDING STATEMENT

This work was supported by the Medical Research Council (MRC) and the University of Bristol MRC Integrative Epidemiology Unit (MC_UU_00011/1, MC_UU_00011/4). CJB is supported by a Wellcome Trust PhD studentship (218495/Z/19/Z). RC is supported by the Research Council of Norway (288083). NMD is supported by the Research Council of Norway (295989). For the purpose of Open Access, the author has applied a CC BY public copyright licence to any Author Accepted Manuscript version arising from this submission. No funding body has influenced data collection, analysis or its interpretation. This publication is the work of the authors, who serve as the guarantors for the contents of this paper.

FIGURE LEGENDS

Figure B1: IBD, left: On the left, the two offspring (red and blue circles) have inherited allele A from their mother (square). This means they are both IBS (i.e., they have the same genotype) and are also IBD (i.e., they both inherited A from their mother).

IBS, right: On the right, both offspring have inherited an A allele. However, the red circle inherited A allele from their mother (square) and the blue circle inherited the A allele from their father (triangle). Thus, the siblings on the left are IBS only (i.e., they have the same genotype but have inherited the alleles from different parents).

Figure 2: The relationship of interest between the exposure (X) and outcome (Y) is demonstrated by the dashed line. The directional arrows from the variable (U) to both the exposure and outcome demonstrate confounding, as U is a common cause of both the exposure and outcome. Therefore, estimates of the association between the exposure and outcome are biased due to the unaccounted presence of the confounding variable.

Figure 3: We demonstrate the indirect maternal genetic effects between mother and offspring, (maternal phenotype to offspring phenotype, red arrow), direct maternal genetic effects (dark blue arrow) and the direct genetic offspring effect on itself (light blue arrow).

Note these effects also hold for the paternal genotype, alongside effects resulting from the likely parental genotype correlation.